# Using Multiple Dermoscopic Photographs of One Lesion Improves Melanoma Classification via Deep Learning: A Prognostic Diagnostic Accuracy Study

Achim Hekler, MSc[1*], Roman C. Maron, MSc[1*], Sarah Haggenmüller, MSc[1], Max Schmitt, MSc[1], Christoph Wies, MSc[1], Jochen S. Utikal, MD[2,3,4], Friedegund Meier, MD[5], Sarah Hobelsberger, MD[5], Frank F. Gellrich, MD[5], Mildred Sergon, MD[5], Axel Hauschild, MD[6], Lars E. French, MD[7,8], Lucie Heinzerling, MD[7,10], Justin G. Schlager, MD[7], Kamran Ghoreschi, MD[9], Max Schlaak, MD[9], Franz J. Hilke, PhD[9], Gabriela Poch, MD[9], Sören Korsing, MD[9], Carola Berking, MD[10], Markus V. Heppt, MD[10], Michael Erdmann, MD[10], Sebastian Haferkamp, MD[11], Konstantin Drexler, MD[11], Dirk Schadendorf, MD[12], Wiebke Sondermann, MD[12], Matthias Goebeler, MD[13], Bastian Schilling, MD[13], Jakob N. Kather, MD[14], Eva Krieghoff-Henning, PhD[1], Titus J. Brinker, MD[1]

1. Digital Biomarkers for Oncology Group, German Cancer Research Center (DKFZ), Heidelberg, Germany
2. Department of Dermatology, Venereology and Allergology, University Medical Center Mannheim, Ruprecht-Karl University of Heidelberg, Mannheim, Germany
3. Skin Cancer Unit, German Cancer Research Center (DKFZ), Heidelberg, Germany
4. DKFZ Hector Cancer Institute at the University Medical Center Mannheim, Mannheim, Germany
5. Skin Cancer Center at the University Cancer Center and National Center for Tumor Diseases Dresden, Department of Dermatology, University Hospital Carl Gustav Carus, Technische Universität Dresden, Germany
6. Department of Dermatology, University Hospital (UKSH), Kiel, Germany

7. Department of Dermatology and Allergy, University Hospital, LMU Munich, Munich, Germany

8. Dr. Phillip Frost Department of Dermatology and Cutaneous Surgery, University of Miami, Miller School of Medicine, Miami, FL, USA

9. Department of Dermatology, Venereology and Allergology, Charité – Universitätsmedizin Berlin, Corporate member of Freie Universität Berlin and Humboldt-Universität zu Berlin, Berlin, Germany

10. Department of Dermatology, University Hospital Erlangen, Comprehensive Cancer Center Erlangen – European Metropolitan Region Nürnberg, CCC Alliance WERA, Erlangen, Germany

11. Department of Dermatology, University Hospital Regensburg, Regensburg, Germany

12. Department of Dermatology, Venereology and Allergology, University Hospital Essen, Essen, Germany

13. Department of Dermatology, Venereology and Allergology, University Hospital Würzburg and National Center for Tumor Diseases (NCT) WERA Würzburg, Germany

14. Department of Medicine III, University Hospital RWTH Aachen, Aachen, Germany

**\*These authors contributed equally to this work.**

**Corresponding author:**

Dr. med. Titus J. Brinker, MD

Digital Biomarkers for Oncology Group, German Cancer Research Center (DKFZ)

Im Neuenheimer Feld 280

69120 Heidelberg, Germany

Phone: +496221 3219304

Email: titus.brinker@dkfz.de

# Abstract

**Background:** Convolutional neural network (CNN)-based melanoma classifiers face several challenges that limit their usefulness in clinical practice.

**Objective:** To investigate the impact of multiple real-world dermoscopic views of a single lesion of interest on a CNN-based melanoma classifier.

**Methods:** This study evaluated 656 suspected melanoma lesions. Classifier performance was measured using area under the receiver operating characteristic curve (AUROC), expected calibration error (ECE) and maximum confidence change (MCC) for (I) a single-view scenario, (II) a multiview scenario using multiple artificially modified images per lesion and (III) a multiview scenario with multiple real-world images per lesion.

**Results:** The multiview approach with real-world images significantly increased the AUROC from 0.905 (95% CI, 0.879-0.929) in the single-view approach to 0.930 (95% CI, 0.909-0.951). ECE and MCC also improved significantly from 0.131 (95% CI, 0.105-0.159) to 0.072 (95% CI: 0.052-0.093) and from 0.149 (95% CI, 0.125-0.171) to 0.115 (95% CI: 0.099-0.131), respectively. Comparing multiview real-world to artificially modified images showed comparable diagnostic accuracy and uncertainty estimation, but significantly worse robustness for the latter.

**Conclusion:** Using multiple real-world images is an inexpensive method to positively impact the performance of a CNN-based melanoma classifier.

# Introduction

Recent projections indicate a substantial increase in global melanoma incidence by 2040, with estimates suggesting increases of up to 50% in cases and 68% in associated death rates.[1] This emphasizes the urgent need for more accurate and efficient diagnostic tools to facilitate early melanoma detection and treatment.

Recent advances in artificial intelligence (AI), particularly in deep learning and convolutional neural networks (CNNs), show promise in assisting clinicians with melanocytic lesion diagnosis. Several studies have demonstrated that CNN-based algorithms perform on par or even surpass expert dermatologists in experimental settings.[2–5] However, good diagnostic accuracy is just one of many aspects that are required for the successful clinical. CNNs still face limitations such as robustness[6,7] and uncertainty issues,[8,9] affecting their reliability.

Robustness refers to how well CNN-based melanoma classifiers maintain accuracy and reliability when input data undergoes alterations like changes in image orientation, lighting and position. Previous studies have shown that when the input images are subject to artificial modifications, the performance of AI-based melanoma detection systems is affected.[10,11] However, such changes are inevitable in routine clinical practice, e.g., there is no standardized orientation for photographing a skin lesion. Hence, it is crucial for CNN-based melanoma classifier to remain stable and accurate under such conditions.

Additionally, the model should effectively communicate the confidence level of its predictions. Model outputs should not be interpreted as probabilities since modern neural networks typically overestimate their own uncertainty, resulting in incorrect predictions made with high confidence.[8] Whenever a model generates an uncertain prediction, it is important that dermatologists are made aware of this uncertainty, allowing them to interpret the results with appropriate caution. Proper confidence estimates are critical in fostering trust between clinicians and AI-based diagnostic tools, as it encourages a collaborative decision-making process that leverages both human expertise and AI capabilities.[12]

To address these limitations, a common technique is to incorporate multiple artificially generated views of the same lesion into the decision-making process. This involves digitally transforming the original lesion image through standard computer vision operations such as rotation, translation, or brightness variations, and averaging the predictions from these transformed images. This technique, known as test-time augmentation, has shown improvements not only in diagnostic accuracy but also in uncertainty estimation and robustness.[10,13,14] However, challenges related to overconfidence and robustness persist.

In this study, we aim to investigate if incorporating multiple real-world images of a lesion, rather than artificially modified images, can enhance this technique. The rationale behind this approach is that real-world images captured from multiple perspectives provide a more comprehensive and nuanced understanding of lesion morphology, reducing the influence of imaging artifacts, such as reflections and partial occlusions. To ensure reliable findings, we utilize data from a prospective multicenter study involving eight university hospitals while focusing on multiple clinically relevant endpoints (i.e., diagnostic accuracy, uncertainty estimation and robustness).

# Methods

## Study design

This prospective, multicenter study was approved by the respective institutional review boards of the participating hospitals/centers and adhered to the Declaration of Helsinki guidelines. STARD 2015 reporting standards were followed and written informed consent was obtained from all participating patients.[15]

Dermoscopic images and patient metadata (e.g., age, Fitzpatrick skin type, lesion localization and diameter) of clinically suspected melanoma were prospectively collected from eight university hospitals in Germany (Berlin, Dresden, Erlangen, Essen, Mannheim, Munich, Regensburg, Wuerzburg) between April 2021 and October 2022 during routine clinical care. For each lesion, a dermatologist captured six dermoscopic images during clinical examination while randomly varying the orientation/angle, position and mode of the dermatoscope (i.e., polarized or non-polarized). To minimize the effect of confounding factors, dermatologists were instructed to avoid well known artifacts (e.g., skin markings). All images were obtained using one of four hardware settings that were available across the participating centers (see **Supplementary Methods**). This dataset will be referred to as SCP2 in the following text.

Subsequently, we trained a binary melanoma-nevus classifier on publicly available dermoscopic images and evaluated its performance on the external SCP2 dataset with respect to three clinically relevant endpoints: diagnostic accuracy, uncertainty estimation and robustness. For model prediction, we evaluated three different methods. The first method, called Single-View, represents the baseline scenario in which only one "original" image per lesion is available and the prediction is performed from that image. For the second method, referred to as multiview-artificial (MV-Artificial), the "original" image is accompanied by artificially modified duplicates generated by applying various image processing techniques such as rotation, zoom and brightness to the "original" image (see **Supplementary Methods**). For the final method, referred to as multiview-real (MV-Real), the "original" image is accompanied by multiple real-world images (i.e., photographs taken in the clinic). At test-time, the model therefore provides a prediction for every single image, which are subsequently combined into an overall prediction (see **Figure 1**).

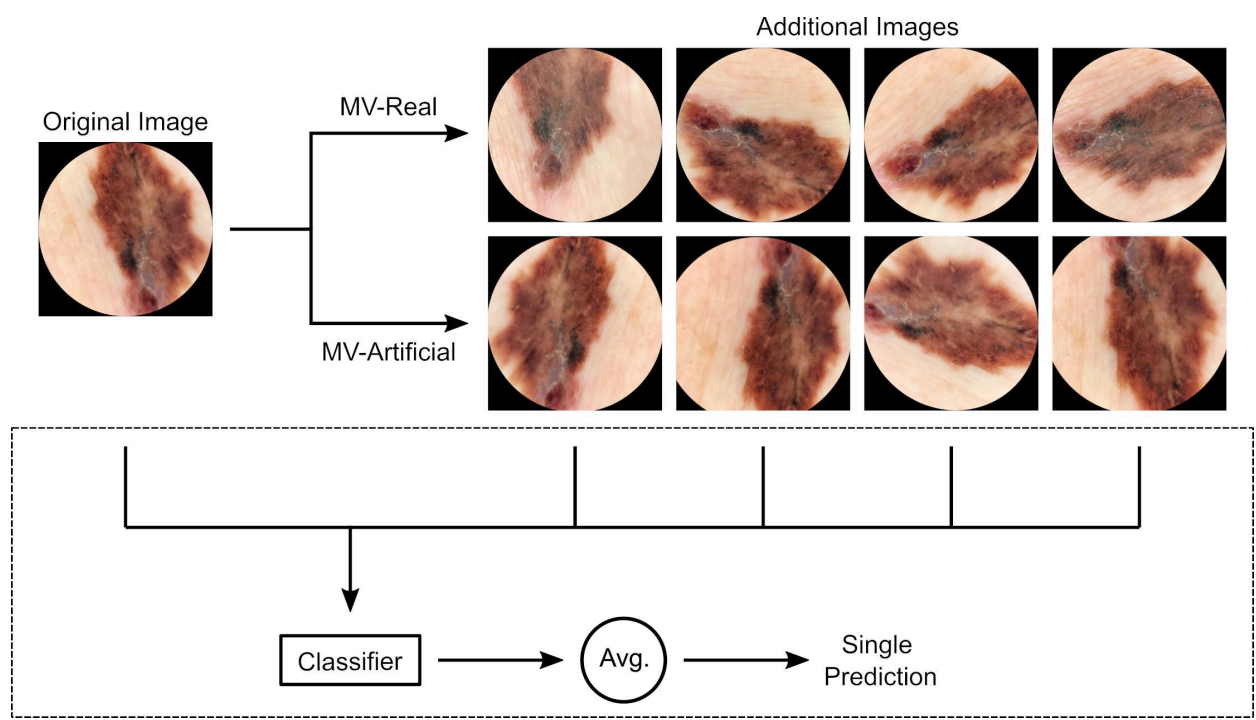

**Figure 1. Illustration of the multiview approach.** Top) For both the MV-Artificial and MV-Real methods, the model makes its final classification based on one original image accompanied by additional lesion images. For MV-Real, the additional images are actual dermoscopic photographs taken in the clinic by the physician (top row). For MV-Artificial, the additional images are artificially created from the original image by applying various image processing techniques such as rotation, zoom and brightness. Bottom) For both approaches, the classifier makes a prediction for the original and each of the additional images. All predictions are subsequently averaged into a single prediction. MV-Artificial: multiview-artificial, MV-Real: multiview-real.

The setup described above is feasible because the SCP2 dataset contains six real-world images per lesion. To ensure that the comparisons and statistical tests in this study were based on the same test data for all three methods, we randomly sampled one image per lesion and labeled this image as the "original" image (referred to as downsampling step). The remaining five images were set aside. Thus, all three prediction methods were evaluated on the same test set, with each image corresponding to a unique lesion. During test-time, the Single-View method received no further images, while MV-Artificial and MV-Real each received five additional images (modified duplicates and real-world images, respectively).

## Participants

Participants were required to be at least 18 years old and have melanoma-suspicious skin lesions that were excised following dermoscopic examination. The suspicious lesions should not have been previously pre-biopsied nor located near the eye or under the fingernails or toenails. Additionally, due to data privacy concerns, lesions with person-identifying features (e.g., tattoos) in their immediate vicinity were excluded from the study. All lesions were histopathologically confirmed by at least one reference pathologist at the corresponding clinic as part of routine clinical practice. In the end, only histopathologically verified melanoma or nevus lesions, recorded until October 2022, were included in this study.

## Model training and evaluation

We trained a CNN with a state-of-the-art ConvNeXT architecture with publicly available melanoma and nevus images from two well-established datasets, HAM10000[16] and BCN20000[17] (see **Supplementary Methods**). At test-time, the Single-View, MV-Artificial and MV-Real approaches were used on the trained model, using the external SCP2 dataset for evaluation. Both training and inference were implemented using PyTorch 1.10.1,[18] CUDA 11.0 and fastai 2.7.10.[19]

## Statistical analysis

The performance of our classifier was evaluated based on three endpoints: diagnostic accuracy, uncertainty estimation and robustness. Diagnostic accuracy was measured using the area under the receiver operating characteristic curve (AUROC), while uncertainty estimation was quantified by the expected calibration error (ECE).[20] The ECE assesses the calibration of predicted probabilities against observed outcomes, with lower values indicating better calibration.

Robustness was evaluated by analyzing the consistency of the classifier's predictions across a series of images per lesion, detecting fluctuations in the model's diagnosis (see **Supplementary Figure 1**). We therefore computed the mean maximum confidence change (MCC), which measures the difference between the model's highest and lowest confidence scores for a series of images. Larger MMC values are worse, as the model's predictions are less consistent. As analyzing robustness requires a series of images per lesion across which to measure fluctuations, we constructed image series of either two or three images per lesion, by using the five additional images which were previously set aside during the downsampling step (see **Study Design**). However, this meant we also had to reduce the number of images used for MV-Real to three and two images respectively. To keep the comparison fair, MV-Artificial was adjusted accordingly.

To reduce the impact of stochastic events, mean values for each metric were calculated using 1000 bootstrap iterations on our test sets. The corresponding 95% confidence intervals (CIs) were determined using the non-parametric percentile method. Statistical testing was conducted for all three hypothesis to identify significant differences between results with our proposed technique (i.e., MV-Real) and those with either the baseline (i.e., Single-View) or the traditional multiview technique (i.e., MV-Artificial). For each endpoint, pairwise Wilcoxon signed-rank tests were used to compare the respective metrics. Significance levels of p<0.05 were adjusted to 0.025 according to the Bonferroni correction (m=2) which equals the expected false discovery rate. In addition, we repeated the downsampling step (see **Study Design**) and all subsequent analysis steps five times in order to ensure that our findings were not based on an unfavorable sample. Statistical analysis was performed using SciPy 1.7.1.

# Results

## Patient characteristics

A total of 617 patients with 656 skin lesions clinically suspected to be melanoma were enrolled in this study. The patient characteristics of the study samples are summarized in **Table 1**. Of the participants, 44.6% were female. The patients' ages at diagnosis ranged from 18 to 95 years, with a median age of 61 years. The distribution of Fitzpatrick skin types was as follows: Type I (8.8%), type II (60.0%), type III (26.4%), type IV (1.3%) and unknown type (3.5%). For 39 patients (6.3%), two different melanoma-suspicious lesions were included in the study, resulting in a total of 656 unique lesions.

**Table 1. Patient characteristics of the study sample.** Distributions of the age at diagnosis, lesion location and lesion diameter are reported.

|  | Melanoma[a] | Nevus |
|---|---|---|
| Patient age at diagnosis (in years) | n=293 | n=363 |
| <35 | 7 (2.4%) | 82 (22.6%) |
| 35-54 | 47 (16.0%) | 124 (34.2%) |
| 55-74 | 124 (42.3%) | 105 (28.9%) |
| >74 | 115 (39.2%) | 52 (14.3%) |
| Lesion location |  |  |
| Palms/soles | 7 (2.4%) | 11 (3.0%) |
| Face/scalp/neck | 65 (22.2%) | 22 (6.1%) |
| Upper extremities | 54 (18.4%) | 36 (9.9%) |
| Lower extremities | 52 (17.7%) | 83 (22.9%) |
| Back | 72 (24.6%) | 120 (33.1%) |
| Abdomen | 17 (5.8%) | 37 (10.2%) |
| Chest | 20 (6.8%) | 40 (11.0%) |
| Buttocks | 2 (0.7%) | 9 (2.5%) |
| Genitalia | 2 (0.7%) | 4 (1.1%) |
| Unknown | 2 (0.7%) | 1 (0.3%) |
| Lesion diameter (in mm) | | |
| ≤ 3.00 | 11 (3.8%) | 63 (17.4%) |
| 3.01 to 6.00 | 27 (9.2%) | 137 (37.7%) |
| 6.01 to 9.00 | 26 (8.9%) | 66 (18.2%) |
| 9.01 to 12.00 | 60 (20.5%) | 52 (14.3%) |
| 12.01 to 15.00 | 46 (15.7%) | 18 (5.0%) |
| > 15.00 | 123 (42.0%) | 27 (7.4%) |

a. Including in situ tumors.

## MV-Real improves diagnostic accuracy and uncertainty estimation compared to Single-View and MV-Artificial

To determine the performance impact of increasing the number of images per lesion, we evaluated our model using three approaches: Single-View, MV-Artificial and MV-Real. Our findings show that MV-Real improves both the diagnostic accuracy and the uncertainty estimation when compared to the Single-View approach. The AUROC significantly increases from 0.905 (95% CI, 0.879-0.929) to 0.930 (95% CI, 0.909-0.951; p<0.001) with the ECE significantly decreasing from 0.131 (95% CI, 0.105-0.159) to 0.072 (95% CI: 0.052-0.093; p<0.001, see **Figure**

**2**). These findings were consistent across all five repeated down-samplings (see **Supplementary Tables 1a-1e**).

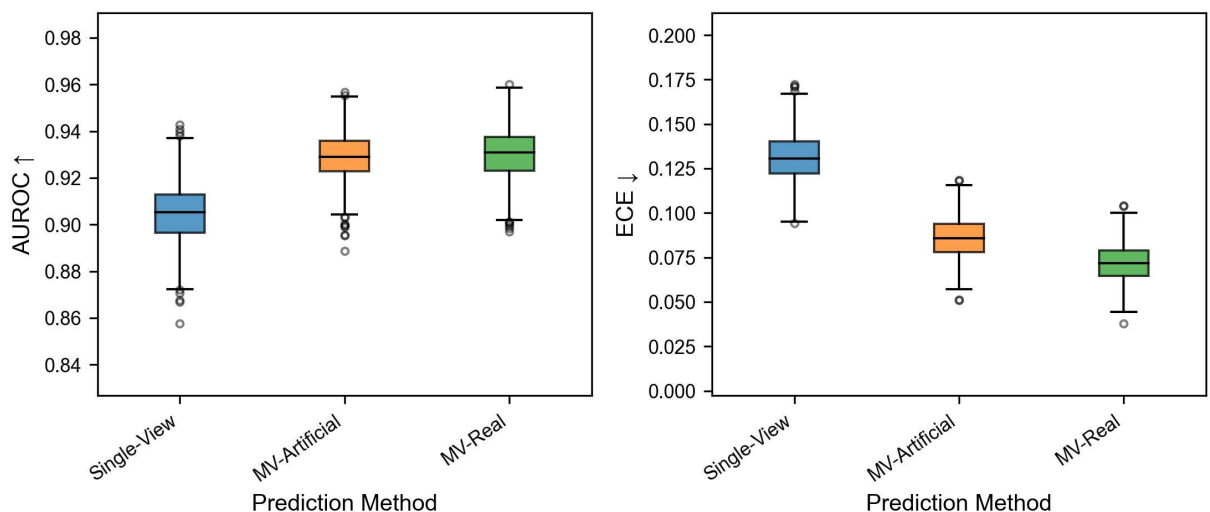

**Figure 2. MV-Real outperforms both Single-View and MV-Artificial with respect to diagnostic accuracy and uncertainty estimation.** The AUROC (diagnostic accuracy) and ECE (uncertainty estimation) are plotted for the three investigated methods on the left and right, respectively. Each box extends from the lower to the upper quartile of the 1000 bootstrap iterations, with a line at the median. In addition, whiskers and fliers indicate the range and any outliers. AUROC: area under the receiver operating characteristic curve, ECE: expected calibration error, MV-Artificial: multiview-artificial, MV-Real: multiview-real.

Similarly, MV-Real also outperformed MV-Artificial which had a significantly lower AUROC of 0.929 (95% CI: 0.908-0.948; p<0.001) and significantly lower ECE of 0.086 (95% CI: 0.064-0.110; p<0.001). These findings were only somewhat consistent across the five repeated down-samplings, with diagnostic accuracy sometimes being on-par or slightly better for MV-Artificial, indicating that there is no practical difference in diagnostic accuracy for both approaches (see **Supplementary Tables 1a-1e**)

## MV-Real improves robustness compared to Single-View and MV-Artificial

The robustness of our model was analyzed across a series of either two or three images per lesion. For the series of three images, the robustness with MV-Real improved substantially over that with Single-View, as the MCC significantly decreased from 0.149 (95% CI, 0.125-0.171) to 0.115 (95% CI: 0.099-0.131; p<0.001), respectively. Similarly, robustness also improved across a series of two images, as the MMC significantly decreased from 0.094 (95% CI, 0.077-0.112) for single-view to 0.066 (95% CI: 0.056-0.076; p<0.001) for MV-Real. Surprisingly, the MV-Artificial method resulted in no robustness improvement at all, having greater MMC values than the Single-View and MV-Real approaches (see **Figure 3**). These findings were consistent across all five repeated down-samplings (see **Supplementary Tables 1a-1e**).

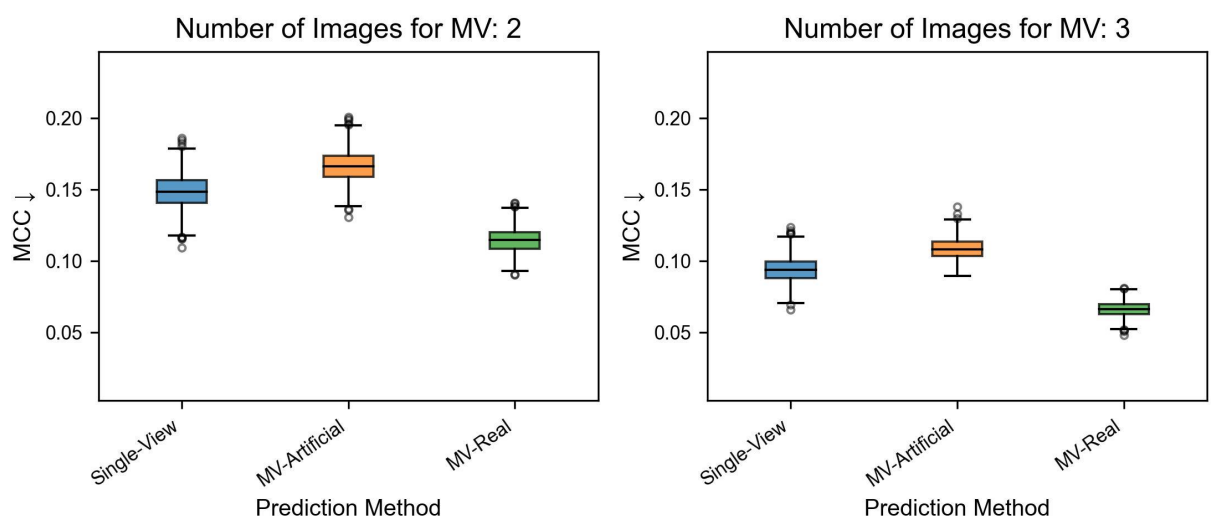

**Figure 3. MV-Real outperforms both Single-View and MV-Artificial with respect to robustness.** Robustness was measured by the maximum change in the classifier's confidence (MCC) across a series of either three (left) or two (right) images. Each box extends from the lower to the upper quartile of the 1000 bootstrap iterations, with a line at the median. In addition, whiskers and fliers indicate the range and any outliers. AUROC: area under the receiver operating characteristic curve, ECE: expected calibration error, MV: multiview, MV-Artificial: multiview-artificial, MV-Real: multiview-real.

# Discussion

## Principal findings

Traditionally, AI-based model make predictions on a single input image, but this approach has limitations due to the brittleness of AI.[10,21,22] Supplying additional images at test-time and subsequently combining the model's predictions into one is an easy-to-implement and already established technique in machine learning.[23] This method – commonly referred to as test-time augmentation – is built on the idea that multiple different images of the same lesion may provide different perspectives on what the model perceives. However, these additional images are typically artificially generated duplicates, offering no new information.

In this study, we aim to investigate whether the performance of an AI-based melanoma classifier could be improved by using real-world images (i.e., photographs taken in the clinic) instead of artificially generated duplicates. Our evaluation focused on three clinically relevant aspects: diagnostic accuracy, uncertainty estimation and robustness.

Our results demonstrate that supplying additional real-world images at test-time (MV-Real) enhances the classifier's performance compared to the traditional case without additional images (Single-View). This improvement was observed across diagnostic accuracy, uncertainty estimation and robustness. It was expected as MV-Real is a modified version of MV-Artificial, which previously showed improvements in these aspects.[10,13,14] Interestingly, when comparing MV-Artificial to Single-View, MV-Artificial performed better in diagnostic accuracy and uncertainty estimation but worse in robustness. This result is surprising since previous studies have indicated robustness improvements with this technique[10]. However, it appears that in this study these improvements were only observed when the image series used to measure robustness was artificially created and in-distribution. In contrast, when the image series consisted of natural real-world images from an external test set, the improvements were less pronounced or absent. Considering our test set also contains natural image series from external test sets, this could indicate that MV-Artificial has limited generalization capabilities regarding robustness and/or that testing on real-world data is substantially different to simulated environments.

Comparing both multiview options, using real-world images (MV-Real) outperformed artificially generated images (MV-Artificial) significantly in uncertainty estimation and robustness, while their diagnostic accuracy performance was comparable. This performance difference can be attributed to the fact that MV-Real provides the model with genuinely new information rather than just variations of old information. Multiple real-world images allow different angles and parts of the lesion to be captured, which is particularly useful for larger lesions or those that are partially occluded (e.g., by hair) or difficult to photograph (e.g., on the ear). Furthermore, the model's classification is not solely based on the quality of a single image. Multiple real-world images increase the chances of having high-quality images or at least having images that are somewhat complementary to each other. As the six images collected for every lesion were a mixture of polarized and non-polarized dermoscopic images, some of the improvements seen with MV-Real could be attributed to this mixture.

## Feasibility in clinical practice

While incorporating a single additional image into the classification already improves the diagnostic accuracy and uncertainty of the classifier, the benefits of including additional images are even more pronounced (see **Supplementary Figure 2**). However, asking physicians to take multiple photographs for every lesion and patient is time consuming and impractical, therefore future work should look into optimizing this process.

## Limitations

The evaluation of our classifier was limited to a binary setting which does not reflect the clinical reality. However, due to the inclusion criteria of our study (i.e., melanoma suspicious skin lesions), the majority of lesions collected were either melanoma or nevus, leaving a large variety of other diagnostic classes which only had insignificant sample sizes. Therefore, our findings may not translate to a multiclass setting.

# Conclusion

Using multiple real-world images of a lesion improves the overall performance of an AI-based melanoma classifier compared to more traditional approaches. As our proposed approach only requires additional photographs, it is easy-to-implement and cost-effective. We therefore recommend integrating it into future clinical workflows, which make use of AI-based computer vision.

## Abbreviations

AI                artificial intelligence

AUROC        area under the receiver operating characteristic curve

CNN            convolutional neural network

ECE             expected calibration error

MCC            maximum confidence change

MV-Artificial    multiview-artificial

MV-Real        multiview-real

## Funding sources

This study was funded by the Federal Ministry of Health, Berlin, Germany (grant: Skin Classification Project 2; grant holder: Titus J. Brinker, German Cancer Research Center, Heidelberg, Germany). The sponsor had no role in the design and conduct of the study; collection, management, analysis and interpretation of the data; preparation, review, or approval of the manuscript; and decision to submit the manuscript for publication.

## Conflicts of interest

Jochen S. Utikal is on the advisory board or has received honoraria and travel support from Amgen, Bristol Myers Squibb, GSK, Immunocore, LeoPharma, Merck Sharp and Dohme, Novartis, Pierre Fabre, Roche and Sanofi outside the submitted work. Friedegund Meier has received travel support and/or speaker's fees and/or advisor's honoraria by Novartis, Roche, BMS, MSD and Pierre Fabre and research funding from Novartis and Roche. Sarah Hobelsberger reports clinical trial support from Almirall and speaker's honoraria from Almirall, UCB and AbbVie and has received travel support from the following companies: UCB, Janssen

Cilag, Almirall, Novartis, Lilly, LEO Pharma and AbbVie outside the submitted work. Sebastian Haferkamp reports advisory roles for or has received honoraria from Pierre Fabre Pharmaceuticals, Novartis, Roche, BMS, Amgen and MSD outside the submitted work. Konstantin Drexler has received honoraria from Pierre Fabre Pharmaceuticals and Novartis. Axel Hauschild reports clinical trial support, speaker's honoraria, or consultancy fees from the following companies: Agenus, Amgen, BMS, Dermagnostix, Highlight Therapeutics, Immunocore, Incyte, IO Biotech, MerckPfizer, MSD, NercaCare, Novartis, Philogen, Pierre Fabre, Regeneron, Roche, Sanofi-Genzyme, Seagen, Sun Pharma and Xenthera outside the submitted work. Lars E. French is on the advisory board or has received consulting/speaker honoraria from Galderma, Janssen, Leo Pharma, Eli Lilly, Almirall, Union Therapeutics, Regeneron, Novartis, Amgen, AbbVie, UCB, Biotest and InflaRx. Max Schlaak reports advisory roles for Bristol-Myers Squibb, Novartis, MSD, Roche, Pierre Fabre, Kyowa Kirin, Immunocore and Sanofi-Genzyme. Wiebke Sondermann reports grants, speaker's honoraria, or consultancy fees from medi GmbH Bayreuth, AbbVie, Almirall, Amgen, Bristol-Myers Squibb, Celgene, GSK, Janssen, LEO Pharma, Lilly, MSD, Novartis, Pfizer, Roche, Sanofi Genzyme and UCB outside the submitted work. Bastian Schilling reports advisory roles for or has received honoraria from Pierre Fabre Pharmaceuticals, Incyte, Novartis, Roche, BMS and MSD, research funding from BMS, Pierre Fabre Pharmaceuticals and MSD and travel support from Novartis, Roche, BMS, Pierre Fabre Pharmaceuticals and Amgen outside the submitted work. Matthias Goebeler has received speaker's honoraria and/or has served as a consultant and/or member of advisory boards for Almirall, Argenx, Biotest, Eli Lilly, Janssen Cilag, Leo Pharma, Novartis and UCB outside the submitted work. Michael Erdmann declares honoraria and travel support from Bristol-Meyers Squibb, Immunocore and Novartis outside the submitted work. Jakob N. Kather reports consulting services for Owkin, France, Panakeia, UK and DoMore Diagnostics, Norway and has received honoraria for lectures by MSD, Eisai and Fresenius. Titus J. Brinker reports owning a company that develops mobile apps (Smart Health Heidelberg GmbH,

Handschuhsheimer Landstr. 9/1, 69120 Heidelberg). The remaining authors declare that the research was conducted in the absence of any commercial or financial relationships that could be construed as a potential conflict of interest.

# Supplementary Materials

## Supplementary Methods

### Dataset description

The four hardware settings across the clinics were as follows:

- HEINE Delta30 dermatoscope with an Apple iPhone 7

- HEINE DELTAone dermatoscope with an Apple iPhone SE

- HEINE DELTAone dermatoscope with an Apple iPhone8

- HEINE IC1 dermatoscope with an Apple iPhone7

The original images were automatically cropped to exclude large parts of the black image margin which originates from dermoscopy and subsequently resized to 300x300 pixels for model training and inference (see section below).

### Model training and evaluation

We utilized publicly available dermoscopic images from the ISIC archive to train a binary classifier capable of distinguishing melanoma and nevus lesions. We focused on the well-documented HAM10000 and BCN20000 subsets of the ISIC archive, containing a total of 29,562 images (7,794 melanoma and 21,768 nevus images). To optimize the hyperparameters of the training process, we employed a five-fold cross-validation procedure, using 30% of the training data for validation in each fold. The model architecture, the number of training epochs, the image size as well as the learning rate were optimized by maximizing the area under the receiver operating curve using Optuna 2.10.0. After determining the optimal hyperparameters, a final model was trained on all 29,562 images (i.e., inclusion of validation set). Subsequently, the model was evaluated on the external SCP2 dataset containing out-of-distribution images.

## Implementation of MV-Artificial

The MV-Artificial approach requires that the original image is duplicated n times and digitally modified before all images are classified by the model. In our case, the digital modifications consisted of rotation, zoom, changes in brightness and warp. Each of these modifications were applied to an image with a probability of 75%. The strength of each modification varied as we considered five different setups: mild, moderate, strong, severe and extreme. The mild setup was considered the default setup and is simply referred to as MV-Artificial in the main manuscript. We used fastai's built-in test-time augmentation function (TTA) with beta set to None as we wanted an unweighted average of all image predictions. The parameters for each setup are listed in the table below.

| Parameter | Mild | Moderate | Strong | Severe | Extreme |
|---|---|---|---|---|---|
| flip_vert | True | True | True | True | True |
| max_rotate | 90 | 90 | 90 | 90 | 90 |
| max_zoom | 1.1 | 1.2 | 1.3 | 1.4 | 1.5 |
| max_lightning | 0.2 | 0.3 | 0.4 | 0.5 | 0.6 |
| max_warp | 0.2 | 0.3 | 0.4 | 0.5 | 0.6 |
| pad_mode | zeros | zeros | zeros | zeros | zeros |

# Supplementary Results

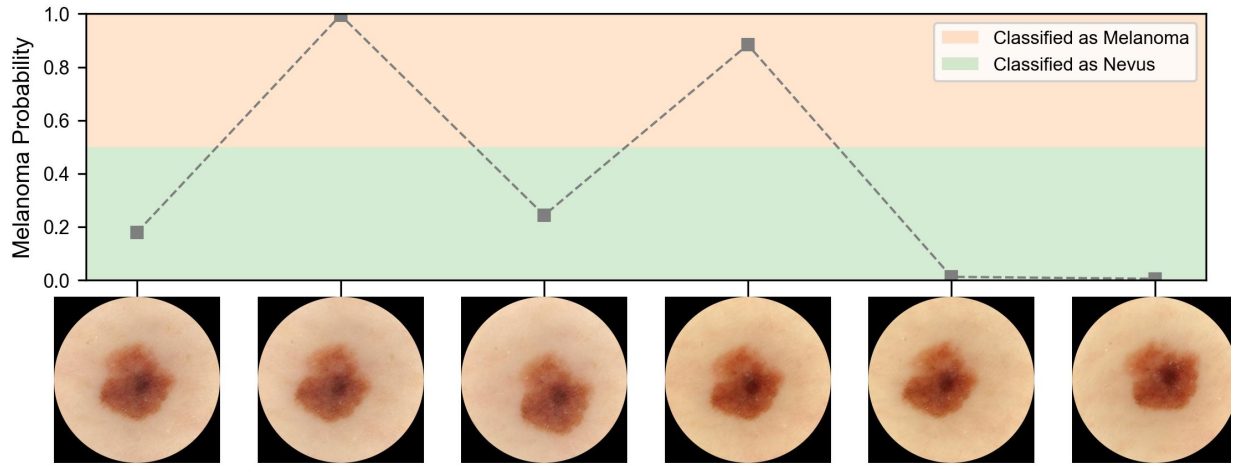

**Supplementary Figure 1. Illustration of how small image changes can cause robustness issues.** The CNN-based algorithm developed in this study classifies multiple images of the same lesion, obtained from our prospective study, as either melanoma or nevus (as indicated by fluctuations in melanoma probability). CNN: convolutional neural network.

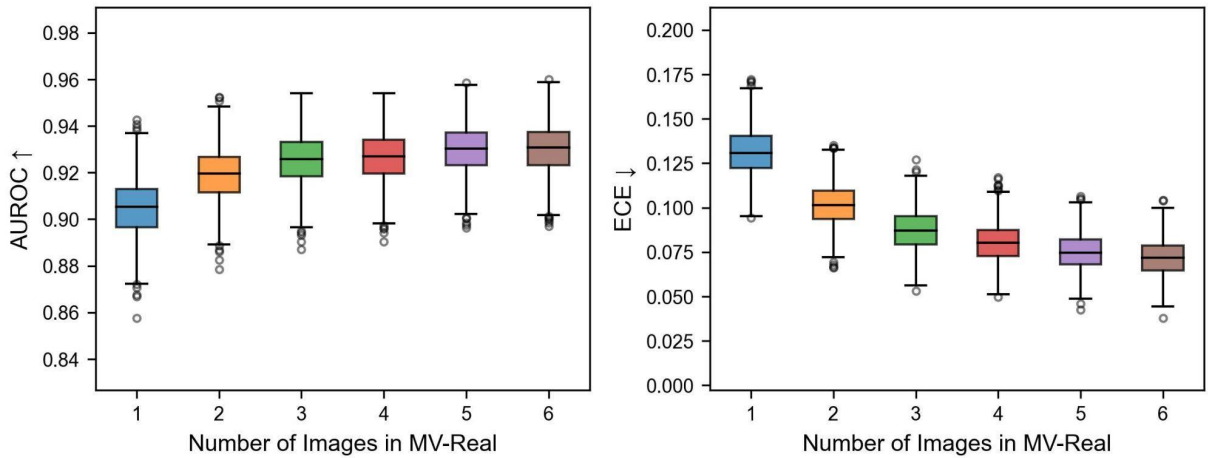

**Supplementary Figure 2. Increasing the number of images used for MV-Real improves diagnostic accuracy and uncertainty estimation.** The AUROC (diagnostic accuracy) and ECE (uncertainty estimation) are plotted for an increasing number of images used during MV-Real. Each box extends from the lower to the upper quartile of the 1000 bootstrap iterations, with a line at the median. In addition, whiskers and fliers indicate the range and any outliers. AUROC: area under the receiver operating characteristic curve, ECE: expected calibration error, MV-Real: multiview-real.

**Supplementary Table 1a. Replication of Results for Different Test Set Samples.**

| Metric | Single-View | MV-Artificial | MV-Real |
|---|---|---|---|
| AUROC ↑ | 0.916 (95% CI: 0.892-0.939) (p<0.001) | **0.930** (95% CI: 0.909-0.948) (p=0.003) | **0.930** (95% CI: 0.909-0.951) |
| ECE ↓ | 0.127 (95% CI: 0.102-0.151) (p<0.001) | 0.087 (95% CI: 0.066-0.110) (p<0.001) | **0.072** (95% CI: 0.052-0.093) |
| MMC (# images: 2) ↓ | 0.145 (95% CI: 0.123-0.168) (p<0.001) | 0.158 (95% CI: 0.137-0.179) (p<0.001) | **0.117** (95% CI: 0.100-0.133) |
| MMC (# images: 3) ↓ | 0.106 (95% CI: 0.087-0.125) (p<0.001) | 0.104 (95% CI: 0.090-0.120) (p<0.001) | **0.069** (95% CI: 0.058-0.080) |

**Supplementary Table 1b. Replication of Results for Different Test Set Samples.**

| Metric | Single-View | MV-Artificial | MV-Real |
|---|---|---|---|
| AUROC ↑ | 0.909 (95% CI: 0.885-0.931) (p<0.001) | 0.927 (95% CI: 0.906-0.946) (p<0.001) | **0.930** (95% CI: 0.909-0.951) |
| ECE ↓ | 0.132 (95% CI: 0.107-0.158) (p<0.001) | 0.087 (95% CI: 0.066-0.109) (p<0.001) | **0.072** (95% CI: 0.052-0.093) |
| MMC (# images: 2) ↓ | 0.149 (95% CI: 0.126-0.173) (p<0.001) | 0.160 (95% CI: 0.140-0.180) (p<0.001) | **0.118** (95% CI: 0.102-0.134) |
| MMC (# images: 3) ↓ | 0.101 (95% CI: 0.083-0.120) (p<0.001) | 0.102 (95% CI: 0.088-0.119) (p<0.001) | **0.069** (95% CI: 0.059-0.081) |

**Supplementary Table 1c. Replication of Results for Different Test Set Samples.**

| Metric | Single-View | MV-Artificial | MV-Real |
|---|---|---|---|
| AUROC ↑ | 0.913 (95% CI: 0.889-0.935) (p<0.001) | **0.933** (95% CI: 0.912-0.950) (p<0.001) | 0.930 (95% CI: 0.909-0.951) |
| ECE ↓ | 0.137 (95% CI: 0.111-0.165) (p<0.001) | 0.083 (95% CI: 0.060-0.107) (p<0.001) | **0.072** (95% CI: 0.052-0.093) |
| MMC (# images: 2) ↓ | 0.152 (95% CI: 0.129-0.175) (p<0.001) | 0.164 (95% CI: 0.143-0.184) (p<0.001) | **0.120** (95% CI: 0.104-0.137) |
| MMC (# images: 3) ↓ | 0.104 (95% CI: 0.085-0.124) (p<0.001) | 0.110 (95% CI: 0.095-0.125) (p<0.001) | **0.076** (95% CI: 0.064-0.087) |

**Supplementary Table 1d. Replication of Results for Different Test Set Samples.**

| Metric | Single-View | MV-Artificial | MV-Real |
|---|---|---|---|
| AUROC ↑ | 0.910 (95% CI: 0.884-0.934) (p<0.001) | **0.931** (95% CI: 0.912-0.949) (p=0.09) | 0.930 (95% CI: 0.909-0.951) |
| ECE ↓ | 0.127 (95% CI: 0.102-0.151) (p<0.001) | 0.086 (95% CI: 0.065-0.108) (p<0.001) | **0.072** (95% CI: 0.052-0.093) |
| MMC (# images: 2) ↓ | 0.143 (95% CI: 0.122-0.166) (p<0.001) | 0.153 (95% CI: 0.134-0.173) (p<0.001) | **0.119** (95% CI: 0.103-0.135) |
| MMC (# images: 3) ↓ | 0.093 (95% CI: 0.076-0.111) (p<0.001) | 0.099 (95% CI: 0.085-0.114) (p<0.001) | **0.068** (95% CI: 0.058-0.078) |

**Supplementary Table 1e. Replication of Results for Different Test Set Samples.**

| Metric | Single-View | MV-Artificial | MV-Real |
|---|---|---|---|
| AUROC ↑ | 0.907 (95% CI: 0.882-0.931) (p<0.001) | **0.931** (95% CI: 0.911-0.949) (p=0.007) | 0.930 (95% CI: 0.909-0.951)) |
| ECE ↓ | 0.132 (95% CI: 0.107-0.157) (p<0.001) | 0.085 (95% CI: 0.064-0.108) (p<0.001) | **0.072** (95% CI: 0.052-0.093) |
| MMC (# images: 2) ↓ | 0.153 (95% CI: 0.131-0.176) (p<0.001) | 0.156 (95% CI: 0.135-0.177) (p<0.001) | **0.116** (95% CI: 0.101-0.133) |
| MMC (# images: 3) ↓ | 0.096 (95% CI: 0.079-0.114) (p<0.001) | 0.101 (95% CI: 0.086-0.117) (p<0.001) | **0.069** (95% CI: 0.059-0.080) |